\documentclass[a4paper,fleqn,usenatbib]{mnras}

\usepackage[T1]{fontenc}
\usepackage{ae,aecompl}

\usepackage{natbib}
\usepackage[toc,page]{appendix}

\usepackage{graphicx}	% Including figure files
\usepackage{amsmath}	% Advanced maths commands
\usepackage{amssymb}	% Extra maths symbols
\usepackage{subcaption}
\usepackage{soul}

\usepackage{setspace}
\title[Chondrule formation]{Protoplanetary cores drove chondrule formation}

\makeatletter
\renewcommand{\@fnsymbol}[1]{\ifcase#1\or *\or **\or \#\or \dagger\or \ddagger\else\@arabic{#1}\fi}
\makeatother

\author[Ali-Dib \& Walton]{
Mohamad Ali-Dib$^{1}$\thanks{malidib@nyu.edu} and  Craig Walton$^{2,3}$\thanks{crw59@cam.ac.uk} \\
$^{1}$Center for Astrophysics and Space Science, New York University Abu Dhabi, PO Box 129188, UAE \\
$^{2}$Department of Earth Sciences, Institute für Geochemie und Petrologie, ETH Zurich, NW D 81.2,\\Clausiusstrasse 25, 8092 Zürich, Switzerland\\
$^{3}$Institute of Astronomy, University of Cambridge, Madingley Road, Cambridge, CB3 OHA, UK
}

\date{Accepted XXX. Received Feb 2025.}

\pubyear{2025}

\begin{document}
\label{firstpage}
\pagerange{\pageref{firstpage}--\pageref{lastpage}}
\maketitle

\begin{abstract}

Chondrules are small spherical objects that formed at high temperatures early in the history of the Solar System. The key compositional characteristics of chondrules may be well explained by high gas pressures in their formation environment \citep{Galy:2000, Alexander:2008}. However, such high gas pressures are widely considered astrophysically unreasonable \citep{Ebel:2023}. Here, we propose that chondrules were formed via the processing of dust grains in the dust-rich envelopes of planetary embryos, {before getting ejected via convective diffusion}. We show that this scenario can explain many salient constraints on chondrule formation, including formation locations; mass and timescale of chondrule production; repeat chondrule heating events; heating timescales; and, most crucially, high prevailing gas pressures. Our work suggests that high gas pressures may indeed have prevailed during the formation of chondrules, reconciling previous analytical observations, experimental evidence, and theory. We suggest that chondrules are mostly the products rather than the precursors of planetary embryo formation -- a result which would have important implications for our understanding of the early history of the Solar System.

\end{abstract}


%%%%%%%%%%%%%%%%%%%%%%%%%%%%%%%%%%%%%%%%%%%%%%%%%%

%%%%%%%%%%%%%%%%% BODY OF PAPER %%%%%%%%%%%%%%%%%%
\section{Introduction}

Chondrules are among the most enigmatic geological materials in our System System. They represent among the oldest and chemically primitive solids known, being younger only than calcium-aluminium-inclusions (CAIs) and some iron meteorites \citep{Connolly:2016}. Chondrules are sub-mm-to-mm-sized spherical objects which appear to have experienced severe and variable thermal histories and display immense diversity of mineralogical compositions and textures \citep{Connolly:1998, Desch:2012}.

The near-Solar compositions of chondritic material and the ancient chronology of their constituent chondrules has been used to argue for a fundamental role of these objects in the early accretionary processes that built the first planetary embryos \citep{Merill:1920}. However, an alternative broad view -- itself encompassing many discrete hypotheses -- posits that chondrules in fact form from processes driven by planetary embryos \citep{Connolly:1998}. These two end-member views have very different implications for our understanding of how planets form and what chondrules can tell us about the early Solar System. However, despite the clear importance of resolving this issue, proposed scenarios for chondrule formation remain controversial.

There are many difficulties in explaining chondrule formation. However, one outstanding issue that applies essentially to all scenarios proposed so far is a mismatch between the high gas pressures that appear to most readily explain chondrule compositions \citep{Galy:2000, Alexander:2008} and the low pressures considered until now to be astrophysically reasonable \citep{Ebel:2023}. 

\begin{figure*}
    \centering
    \includegraphics[width=13cm]{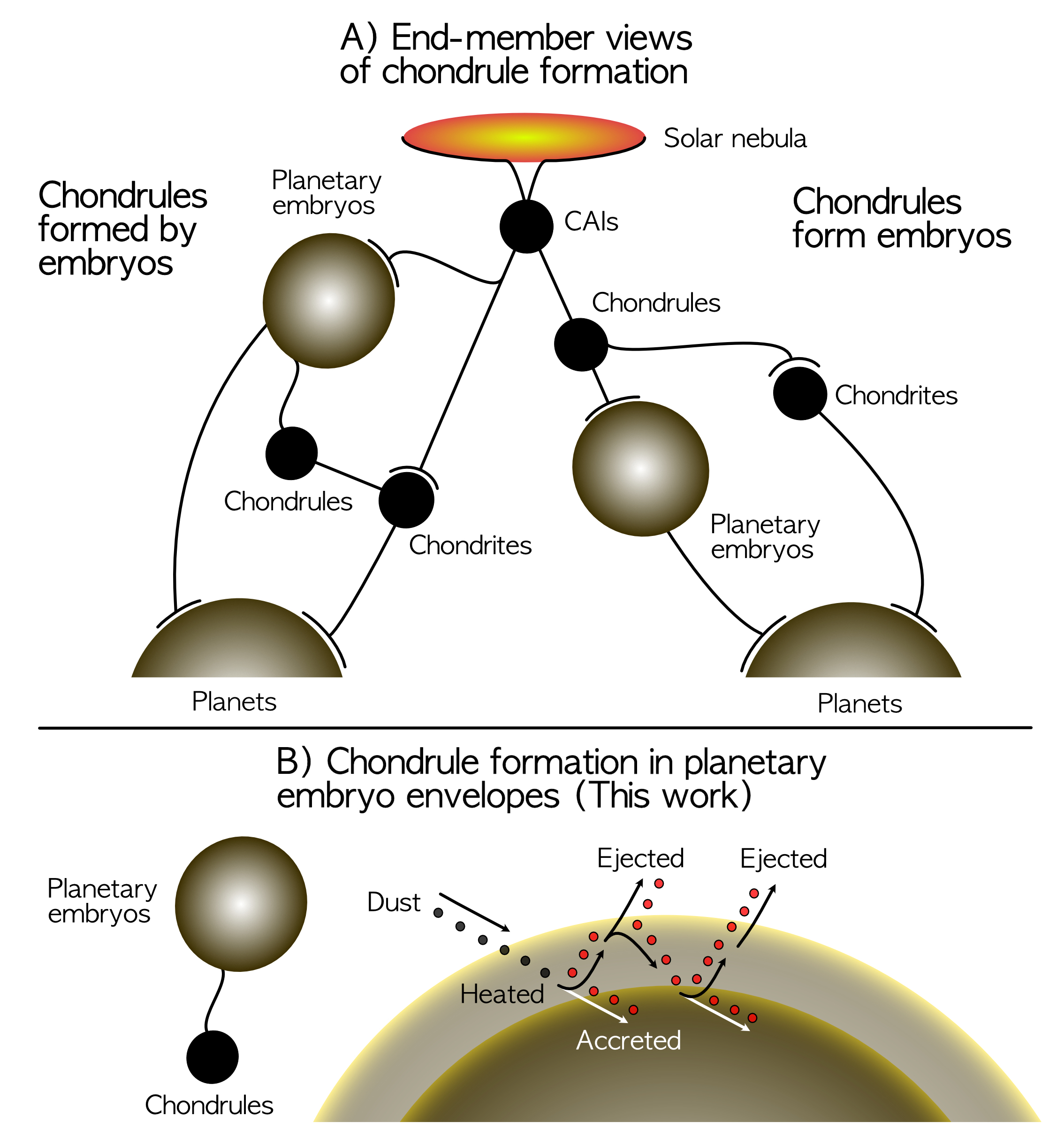}
    \caption{Schematic view of chondrule formation scenarios. A) Two end-member views are shown. On the left, the branch of scenarios is depicted in which chondrules are generally products of planet formation. On the right, the branch of scenarios is depicted in which chondrules are generally precursors to planet formation, i.e., forming earlier than planetary embryos and acting as seed materials for their eventual growth. B) Detailed schematic view of our proposed embryo-first scenario for chondrule formation. We propose that planetary embryos produces chondrules during interactions with dust particles, which enter the atmospheric envelope, become heated, and some fraction of which is then ejected {via convective diffusion through the envelope}. This mechanism allows for the possibility of some small fraction of this dust following a path involving multiple heating cycles.}
    \label{fig:schematic}
\end{figure*}

Here, we propose that chondrules formed via the dynamical interactions of dust grains with the hot, dusty, and hydrogen-rich envelopes of planetary embryos. This scenario, summarized in Fig. \ref{fig:schematic}, uniquely provides high gas pressures during chondrule formation, whilst still performing well when attempting to match other observations of chondrules. %Our work suggests that chondrules are mostly the products rather than the precursors of planetary embryo formation. This result has important implications for the use of chondrules as an archive of early Solar System history and for our understanding of the factors that govern late accretion of volatile elements to terrestrial planets.

\section{Methodology}

\subsection{Atmospheric structure}
Our model and methodology mirrors that of \citep{alidib} to a large degree, with moderate modifications.
We model the atmosphere using the standard atmospheric structure equations. The equation of hydrostatic equilibrium is expressed as:

\begin{equation}
\frac{dP}{dr} = -\frac{GM}{r^2}\rho(r)
\end{equation}

The temperature gradient equation assumes that heat is transported either by radiation (in convectively stable regions) or by adiabatic convection (in unstable regions). It is written as:

\begin{equation}
\frac{dT}{dr} = \nabla \frac{T}{P} \frac{dP}{dr}
\end{equation}

Here, $\nabla$ is defined as $\nabla = \text{min}(\nabla_{\mathrm{ad}}, \nabla_{\mathrm{rad}})$ and follows the Schwarzschild convective stability criterion ($\nabla_{\mathrm{rad}} < \nabla_{\mathrm{ad}}$). The adiabatic gradient, $\nabla_{\mathrm{ad}}$, is given by:

\begin{equation}
\nabla_{\mathrm{ad}} \equiv\left(\frac{d \ln T}{d \ln P}\right)_{\mathrm{ad}} = \frac{\gamma - 1}{\gamma}
\end{equation}

where the adiabatic constant $\gamma = 4/3$. The radiative gradient, $\nabla_{\mathrm{rad}}$, is expressed as:

\begin{equation}
\nabla_{\mathrm{rad}} \equiv \frac{3 \kappa P}{64 \pi G M\sigma T^{4}} L
\end{equation}
where  $\kappa$ is the opacity and $L$ represents the accretion luminosity of the envelope defined as:
\begin{equation}
L=\frac{G M_c \dot{M}_{acc}}{R_c}
\end{equation}
and it is generated by pebble accretion at a rate defined following \citep{lamb} as $\dot{M}_{acc}$:
\begin{equation}
\dot{M}_{{acc}} = 4.8 \times 10^{-6} \left( \frac{M_{\text{core}}}{M_{\oplus}} \right)^{2/3} \left( \frac{R}{10 AU} \right)^{-\frac{5}{12}}
\end{equation}
where $R_c$ is the core radius. We define the total opacity $\kappa$ as:

\begin{equation}
\kappa=\kappa_{\mathrm{gas}}+\kappa_{\mathrm{gr}}
\end{equation}

with the grain opacity following \citep{alidibthompson} :

\begin{equation}
\label{grainopeq}
\kappa_{\mathrm{gr}}=\kappa_{\mathrm{geom}} Q_{e} = \frac{3Z_{gr}}{4\rho_s a}\times min\left(\frac{0.6\pi a}{\lambda_{max}}, 2\right)
\end{equation}

Here, $Z_{gr}$ is the grain abundance, $a_s$ is the grain size, and $\rho_s$ is their internal density. We use $\rho_s = 3 \ \mathrm{g/cm^3}$ for both the planetary core and dust particles internal density.
The gas opacity is defined as a function of metallicity, pressure and temperature following \citep{freedman} and includes $H-$ at high temperatures.

The equilibrium dust size in the envelope $a_s$ is set by two competing processes: grain growth through coagulation \citep{ormel} and grain collisional destruction \citep{alidibthompson}. The relative importance of these two processes is decided mainly by whether the collisional speeds reach the silicate fragmentation threshold ($V_f \sim$ 100 cm/s for silicates and 1000 cm/s for ices \citep{blum}).
Here we define the atmospheric dust size as the smallest of three characteristic sizes: 
growth limited, convective-fragmentation limited, and drift-fragmentation limited size. The fragmentation limits are obtained following \citep{alidibthompson} by setting the collisional speeds equal to the fragmentation speed $V_f$. The growth limited size is obtained following \citep{ormel}. 

%We hence obtain for $a_{s,tf}$:
%\begin{equation}
    %\frac{\rho_s\, a_{s,tf}}{{\rho_g}\cdot c_s}\cdot \max\!\Biggl(\frac{4}%{9},\,\frac{a_{s,tf}}{l_g}\Biggr)
%=\frac{4\cdot 3.14\,(V_f)^2\, r^3\, {\rho_g}}{{L}} 
%\end{equation}
%where the Epstein and Stokes regimes stopping time were folded into the equation, $l_g$ is %the gas mean free path. 
%The equation for $a_{s,df}$ is 
%\begin{equation}
%    \frac{\rho_s\, a_{s,df}}{{\rho_g}\cdot c_s}\cdot \max\!\Biggl(\frac{4}%%{9},\,\frac{a_{s,df}}{l_g}\Biggr)
%=\frac{4\cdot 3.14\,(V_f)^2\, r^3\, {\rho_g}}{{L}} 
%\end{equation}

As discussed in \citep{alidibthompson, alidib}, collisions in these envelopes can be destructive, leading in many cases to small equilibrium dust size and, in consequence, to a high opacity convectively unstable envelope.

Finally we close the system with the ideal gas equation of state $P=\rho_g k_B T/\mu$. We solve these equations by integrating inwards from the outer boundary at R$_{\rm out}$, the minimum of the Bondi and Hill radii, to the core. We assumed the disk is radiative and calculate its temperature and density following \citep{alidibcumming}:
\begin{equation}
T_d = 373 \ r_{au}^{-9/10} K \\ \ \ \mathrm{and} \ \  
\rho_d = 1.7\times 10^{-10} \ r_{au}^{-33/20} g/cm^3
\end{equation}

A crucial unique characteristic of chondrules is their cooling history. We calculate the cooling rate $\Lambda$ of dust particles accounting for radiative and conductive cooling as:

\begin{align}
\label{cle}
    \Lambda = \left(\frac{\epsilon A \sigma_{SB}}{mc}\right)
\left(T_{gr}^4 - T_g^4\right)  \\
+ \left(\frac{h A}{mc}\right)
\left(T_{gr} - T_g\right)
\end{align}
where $A$ and $m$ are the dust particle's surface area and mass, $c$ its specific heat capacity, $\epsilon$ its emissivity, and $h$ the heat conduction coefficient. We use $\epsilon=0.1$, $h=10^5$ erg/cm$^2$sK, and $c=8.36\times 10^6$ erg/gK.

%    \Lambda = \left(\frac{0.3 \cdot \sigma_{SB}}{\rho_s \cdot 8.36 \times 10^6 \cdot %a_{\text{s}}}\right)
%\left(T_{gr}^4 - T_g^4\right)  \\
%+ \left(\frac{3 \cdot 10^5}{\rho_s \cdot 8.36 \times 10^6 \cdot a_{\text{s}}}\right)
%\left(T_{gr} - T_g\right)

\subsection{Priors \& sampling}
\label{samp}
We sample our model's input parameters (r$_{au}$, Z$_{gr}$, M$_{c}$) randomly from log-uniform distributions and generate a total of around 50,000 envelopes. r$_{au}$ is sampled from 0.4 AU (semimajor axis of Mercury) to 30 AU. The atmospheric metallicity Z$_{gr}$ is sampled from strongly subsolar (Z$_{gr}$=10$^{-4}$ $\times$ solar) to supersolar (10 $\times$ solar) to reflect the uncertainty on this parameter. The core mass M$_{c}$ is sampled from 0.1 M$_\oplus$ (atmospheres of smaller cores are too cold) to 10 M$_\oplus$ where gas accretion begins. Finally, we keep only atmospheres that satisfies two a priori conditions necessary for chondrule formation in our scenario:

\begin{enumerate}
\item The envelope is fully convective beyond the dust vaporization point (1500 K) for dust to be able to diffuse from that point outwards toward the surface.
\item The maximum temperature in the envelope is higher than 1500 K to vaporize silicates.
\end{enumerate}

In section \ref{paramspace} we show how these conditions are commonly met throughout the explored input parameters space. Our model can thus form chondrule with no fine tuning necessary.

\section{Results \& discussions}

\begin{figure*}
    \centering
    \includegraphics[width=16cm]{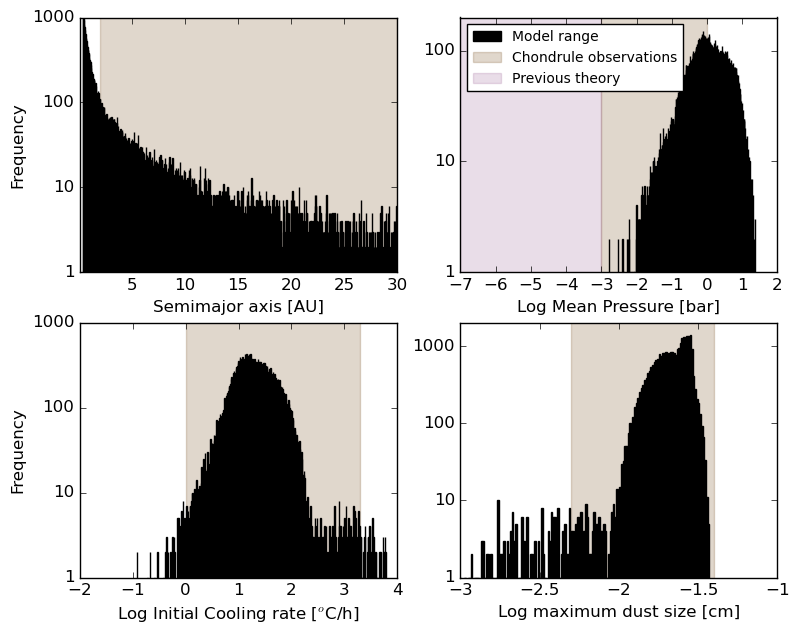}
    \caption{Model output for the conditions experienced by dust grains and their physical parameters in our model. Only scenarios where the atmosphere is hot enough to vaporize silicate and grains can escape the embryo are considered in the data plotted here. A) Top left: our model is agnostic with respect to semimajor axis and can form chondrules throughout the disk. B) Top right: average pressures are many orders of magnitude higher than achieved in previously proposed  scenarios. Our results are strongly consistent with the the combined experimental and modeling evidence of \citep{Galy:2000} that suggested high gas pressures to be most consistent with observations of the chemical composition of chondrules. C) Bottom left: cooling rates are consistent with the literature synthesis of estimated chondrule cooling rates by  \citep{Desch:2012}.  D) Bottom right: ejected particle sizes in the model output overlap the range of observed sizes of chondrules \citep{Friedrich:2022}. Our model is simplified and does not account for particle size evolution during transport. }
    \label{fig:obscomp}
\end{figure*}

\subsection{General overview}
\label{generaloverview}

\subsubsection{Comparison to observational constraints}
Observations that are challenging to simultaneously reconcile in chondrule formation scenarios include evidence for (i) high ambient gas pressure during chondrule formation, (ii) chondrule size distributions, (iii) wide-ranging cooling rates of around 1--to--1000 K/hr \citep{Desch:2012}, (iv) oxygen isotope and nucleosynthetic anomaly evidence for formation across both the inner and outer Solar System \citep{Yap:2023}, (v) the restricted timeframe of the chondrule formation epoch, from around 1 to 3 Myr after CAIs \citep{Connelly:2012, Bollard:2017} and the relatively high abundance of chondritic material in the Solar System; and (vi) rare instances of multiple heating cycles for individual chondrules \citep{Connolly:2016}; We will now argue that chondrule formation in protoplanetary envelopes is a plausible way to explain all of these features. In Figure \ref{fig:obscomp} we show the distribution of the main relevant output quantities in our model, as they compare to observations. 

These include a- the semimajor axis of the cores, b- mean pressure, c- the initial cooling rate of dust particles at 1500 K, and d- the maximum dust size in the envelope. The takeaway from this plot is that our model produces envelopes with enough variety to account for the observational diversity of chondrules. We discuss these quantities in greater details below. 

\subsubsection{Input parameters space}
\label{paramspace}

\begin{figure}
    \centering
    \includegraphics[scale=0.31]{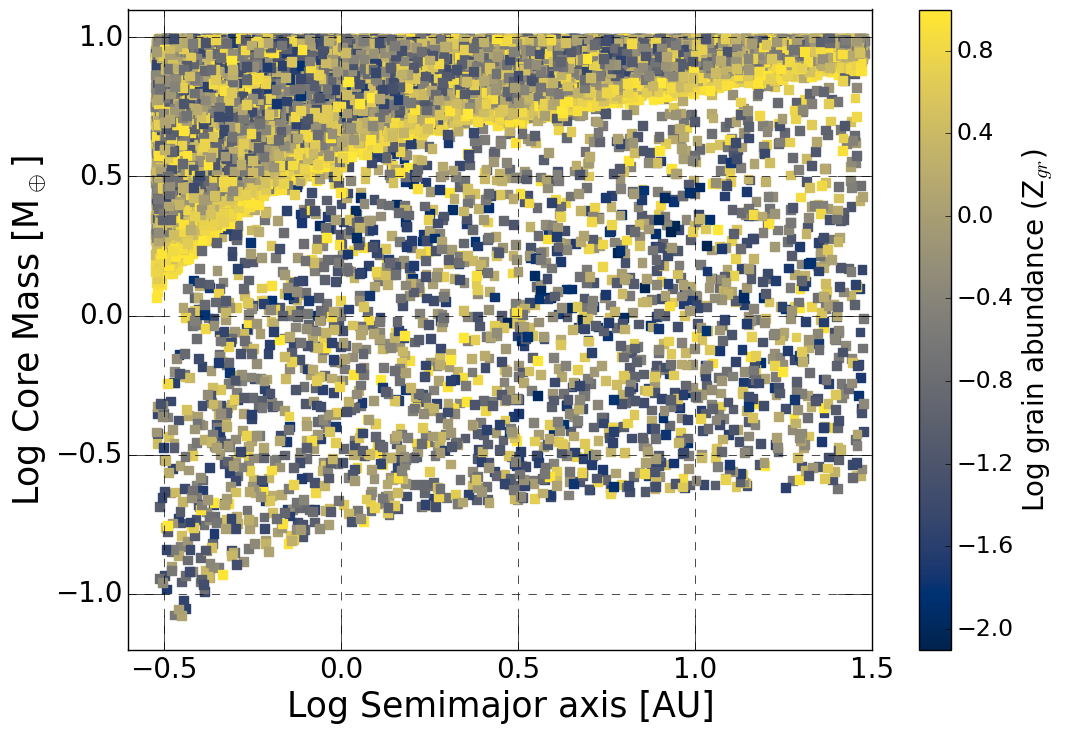}
    \caption{The sets of semimajor axis, core mass, and envelope metallicity that allow for the formation of chondrules following the conditions enumerated in section \ref{samp}. This plot represents an overview of the possible parameters space in our model and highlights its broadness.  }
    \label{fig:main_all}
\end{figure}

In Figure \ref{fig:main_all} we show the values of our input parameters that allow for the formation of chondrules, and account for their observed properties.

%In addition to the conditions of section \ref{samp}, we add :
%\begin{enumerate}
%    \item The maximum pressure in the atmosphere is high enough ($>$ 0.1 bar) for chondrule %formation as discussed above. 
 %   \item The initial cooling rate of the thermally processed dust particles is above 1 %$^o$C/hr as observed.
%\end{enumerate}

Fig. \ref{fig:main_all} shows first that chondrules in our model can form anywhere in the disk, a significant finding consistent with the experimental constrains on chondrules origins. We note that the difference in scatter density between low and high mass cores in the plot is simply due to the atmospheres of massive cores being hotter. Since we are sampling from a uniform distribution, this naturally will lead to higher scatter density for these cores. This is hence a consequence of our priors, and does not imply a higher probability for the formation of chondrules in the envelopes of massive cores. 

Furthermore, our model allows for chondrule formation for a wide range of protoplanetary core masses. These range from {$\sim 1$ Mars mass in the inner most nebula to} 0.5 to 10 $M_\oplus$ throughout the entire disk. Finally, Fig. \ref{fig:main_all} also shows that atmospheric metallicities ranging from sub to supersolar values are allowed. {Condensation sequence calculations favor metallicities 10-1000 the solar value for chondrule formation \citep{ebel}. While such values are allowed by our model, $Z_{gr}$ remains a free input parameter. Detailed modeling of the evolution of $Z_{gr}$ in the envelopes of low mass cores have been done by \cite{alidibthompson}, who found $Z_{gr}$ to increase self-consistently to values approaching 100$\times$ solar.}

\subsection{Heliocentric formation distances}

We start by considering formation distances relative to the Sun. Chondrules have been found in all primitive Solar System materials, including in outer Solar System cometary materials \citep{Bridges:2012}. It has been proposed that this is evidence for transport of chondrules from an inner Solar System formation locus out to the formation locus of outer Solar System primitive materials \cite{Shu:1997, Ciesla:2007}. However, the isotopic composition of such chondrules strongly suggests their formation in situ at a wide range of heliocentric distances \citep{Clayton:1991, Desch:2010}. This observation is consistent with our model, where the only requirement for chondrule production is accretion to and ejection of dust from a hot young planetary embryo (Fig. \ref{fig:main_all}). Our scenario therefore accounts readily for chondrule production throughout the early Solar System, largely removing the challenging requirement for large distances of radial transport post-formation. {We note that, in the innermost part of the protosolar nebula, core masses as low as Mars are enough in our model to form chondrules due to the high ambient pressures and temperatures. Such small cores can readily form even via planetesimal accretion, if indeed pebble accretion does not operate in the inner solar system as suggested by the geochemical arguments of \cite{morby2025}.}

%We find a weak relationship between chondrule production, planetary embryo mass, and heliocentric distance (Fig. X). Embryos forming further from the Sun do produce chondrules at lower masses than those closer in (WHY?), but this is a weak effect. Overall, the most important variable in determining the locus of chondrule formation is simply the formation locations of planetary embryos themselves. Given their expected formation all throughout the early Solar System, and the existence of planets all throughout the Solar System today, this would appear to fit well with observations for variable chondrule formation distances from the Sun. This result again appears to compare favorably to the output of other modeled scenarios (Fig. X). 

\subsection{Pressures}

%Whilst there are several areas in which we find a good fit between our model output and chondrule observations (Fig. 2), the most important is that our scenario for the first time offers the possibility of high ambient gas pressure during chondrule formation.

There has long been experimental evidence that chondrule formation occurred under high gas pressures (perhaps 1 bar or more) \citep{Galy:2000}. {This is around 1,000,000 to 1,000 times higher gas pressure than than is typically achieved in nebular models of chondrule formation \citep{Alexander:2008, pressure1,pressure2,Ebel:2023, pressure3}}

(Fig. \ref{fig:obscomp}), leaving gaps between experiment, theory, and observation. Our work changes this status quo.

Examining the pressure-temperature histories for dust grains ejected from the protoplanetary envelope in our model, we find that the average pressures experienced by chondrules bar range from 0.01 to several bar (Fig. \ref{fig:obscomp}). If the evidence for high gas pressures during chondrule formation is taken at face value, our scenario is the only one so far proposed that yields results within error of the gas pressure constraint (Fig. \ref{fig:obscomp}).

\subsection{Protochondrule sizes}

Size distribution is a key chondrule property. However, the evolution of dust size in the envelopes of protoplanets is a complex problem well beyond the scope of this paper, so our model can only give general insights on the chondrule sizes generated by this mechanism.
In Figure \ref{fig:obscomp} we show how the maximum dust sizes (largest particles) in the envelopes are consistent with observations. Collisional-equilibrium dust sizes, in fully convective atmospheres, tend to increase in the outer atmosphere as they become dominated by convective fragmentation instead of drift-fragmentation, in which case we have $a_{con} \sim r^3$ \citep{alidibthompson}.

\subsection{Total mass and timeframe of chondrule production}

A major constraint on chondrule formation models is the total chondrule mass they produce, where it is expected to be in the order of 0.005 $M\oplus$ or $\sim$10 times the Main Belt's current mass  \citep{Deienno:2024}.

To estimate the total mass ejected by our model, three ingredients are needed: the number of cores in the disk, the accreted mass per core, and the ejected fraction of this mass. 
The total accreted mass is known since it is controlled by the pebble accretion rate of the core, and, over the disk lifetime, is of the same order as the core's mass. The number of cores in the protosolar nebula and their mass distribution is unknown and subject to much debate, so we limit ourselves to a simple nominal case: the cores of the 8 solar system planets, with one extra ice giant. We note that scenarios like oligarchic growth with many more smaller cores in the disk are not relevant here as the core masses needed to create chondrules are higher than that of the majority of early oligarchs.

Finally, we need to estimate the ejection rate of chondrules {(and their overall ejection paths for further study)}. For this we use a simple 1D random-walk scheme where dust particles are injected at the silicate vaporization radius, then diffuse due to convection\footnote{Thus assuming diffusive convection dominated by small scale eddies.} till they either hit the core or get ejected. The step size is sampled from a Gaussian distribution:
\begin{equation}
\Delta x = \sqrt{2 \cdot D_{\text{con}} \cdot \Delta t} \cdot \mathcal{N}(0,1)
\end{equation}
where $D_{\text{con}}$ is the convective diffusion coefficient :
\begin{equation}
    D_{\text{con}} = \left( 1 + \text{St}^2 \right)^{-1} HV_{\text{con}}
\end{equation}
with $\text{St}$ the dust size-dependent Stokes number. 

The ejection rate depends on the mass and location of the core, and thus is controlled by our assumption on number and mass of cores. For our 9 planetary cores scenario, we find $M_e \sim 0.009$ $M\oplus$, which is again consistent with observations. 

%Exploring a range of gas/dust ratios and accretion rates from the literature, we find that the total mass of chondrules produced is within error of the estimated mass of the primordial Main Belt (Fig. X). 

Another scenario which can in principle produce similar quantities of chondrules includes impact splashing \citep{Lichtenberg:2018}. Our model naturally couples the epoch of chondrule formation to the stage of dust accretion to hot planetary embryos, which self-terminates once planets grow and cool and/or during disk dissipation (Fig. 1). The age distribution of chondrules is therefore pre-determined in our scenario to match the observation that chondrule ages are confined to the lifetime of the disk.

In contrast, impact-splash models produce chondrules not continuously but in discrete collisional events. Given that the larger collisions produce exponentially more mass of ejecta, such models can yield the right mass of material but not so easily the right age distribution. Indeed, there are temporally and compositionally anomalous chondrules  that present compelling evidence for having formed in a post-impact vapor plume \citep{Krot:2005}. However, given the anomalous textures and compositions in this case, we argue that the CB chondrules may well be impact-related but that the majority of chondrules may not. Indeed, impact splash scenario may struggle to explain the near-Solar compositions of chondrules \citep{Lichtenberg:2018}, since impacts involving most of the mass at this early time in Solar System history would involve molten (and hence very likely differentiated) objects.

\subsection{Cooling history}

Cooling rates are an area where many scenarios are able to produce reasonable fits and is therefore not strongly discriminatory between the scenarios \citep{Schraeder:2018}. Nonetheless, our model produces a wide variety of heating pathways for dust grains that compares well with reconstructed chondrule cooling rates (Fig. \ref{fig:obscomp}).

\begin{figure*}
    \centering
    \includegraphics[scale=0.7]{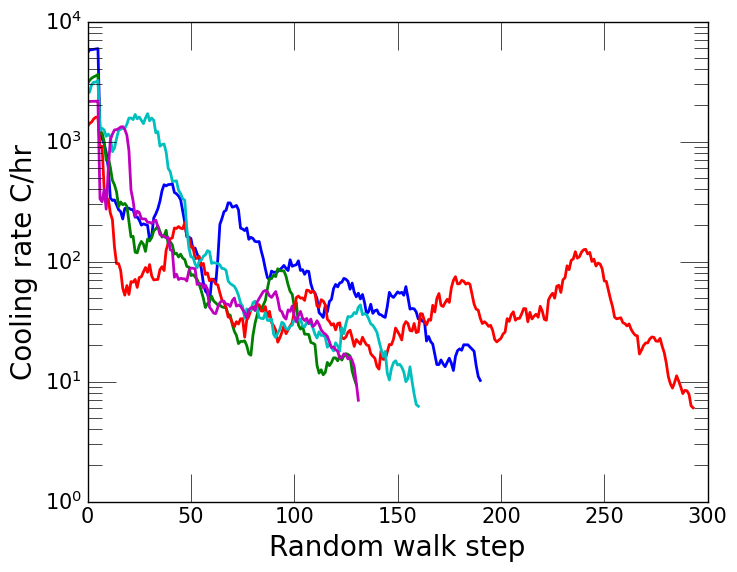}
    \caption{The cooling rate as a function of time (random walk steps) for 5 random particles in our model. The particles are all injected at 1500 K and get eventually ejected from the atmosphere. The range of cooling rates encountered by these particles is consistent with chondrule measurements.  }
    \label{fig:cooling}
\end{figure*}

To get a statistical understanding of these quantities, we couple our diffusion model discussed above to eq. \ref{cle} describing the cooling rate due to conduction and radiation, and we track particles that eventually get ejected through their heating and cooling cycles. In Fig. \ref{fig:cooling} we show example cooling rate evolution for random particles as they diffuse throughout the atmosphere of a 1 $M_\oplus$ protoplanet at 1 AU, from the dust vaporization radius at 1500 K out to the disk. These protochondrules encounter a variety of cooling times throughout this evolution, but will spend the majority of the time in the 10-1000 $^o$C/hr zone, consistent with observations. This plot shows moreover that our model is consistent with the short duration (no longer than minutes to hours) for chondrule heating at the peak temperature \citep{connolly}, as every random walk step corresponds to $\Delta t = $1000 s. 

Note that in this plot we show only when the cooling rate is positive, but in reality, due to the randomness of the process, the particles go through intermittent heating periods as well as they get kicked inwards. These can be ignored as long as they happen far outside the silicate vaporization radius in cool regions.

\begin{figure*}
    \centering
    \includegraphics[scale=0.7]{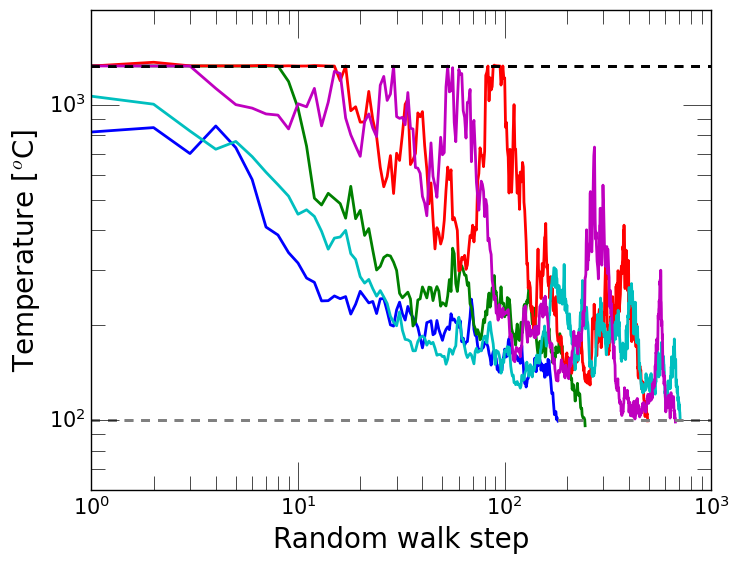}
    \caption{The temperatures encountered by diffusing dust particles as a function of time (random walk steps) for 5 random cases in our model. While most particles are heated only once, many encounter the silicate vaporization line multiple times and thus are consistent with multi heating-events chondrules.}
    \label{fig:ejection}
\end{figure*}

There is evidence however that a small percentage of chondrules suffered repeat heating cycles \citep{connolly}. Our proposed scenario naturally includes mechanisms to give rise to repeated heating cycles of dust grains, whereby grains may fail to escape the embryo envelope, sink, heat up again, and then subsequently be ejected.
In Fig. \ref{fig:ejection} we show the temperature (thus the radial) trajectory of particles on their way to being ejected from the envelope. Here we find chondrule precursors that underwent both single and multiple heating cycles throughout their evolution. While the temperature of all particles fluctuates back and forth, it is for only a minority of particles where these stochastic fluctuations happen in the vicinity of the vaporization line at 1500 K. It is only this minority hence that will exhibit multi-heating cycles characteristics. For the same nominal example, we find the multi-to-single stage heating ratio to be around 20\%. This compares very well against observations where 15\% of all Fe- and Mg-rich chondrules seem to have been repetitively heated \citep{connolly}.

\subsection{Chemical composition}

Finally, we speculate that a planetary embryo formation environment for chondrules may help to explain the incredible compositional diversity of chondrules that formed at the same time and at the same heliocentric distances. Currently, it is unclear what mechanisms could fractionate C, O, and H in order to produce the enormous variations in oxygen fugacity that seem needed to explain the compositions of chondrules \citep{fug1,fug3,fug2}. However, recent evidence suggests that significant chemical gradients existed in the atmospheres of planetary embryos \citep{fegley,chem2,madhu}. The mass of the core, its surface composition, collisional and migration histories, and time of formation all play a role in controlling the chemical composition of planetary atmospheres. For example, it is easier to saturate with water vapor the outer atmosphere of a lower core mass, leading to a wet adiabat, and thus completely different atmospheric structure than a slightly more massive core at the same location \citep{alidibcoll}. Therefore, we suggest that dust grain entry into, descent through, and emergence from either chemically evolving or stratified protoplanetary  envelopes or atmospheres with different chemical composition may help to explain these confounding observations. Correlations between depth of particle descent, pressure-temperature pathways, and the composition of envelope layers encountered may in future provide extremely specific predictions with which to compare against chondrule observations.

\section{Implications, caveats, and future work}

It is beyond the scope of this work to deal with the chemical pathways involved in chondrule formation. Future work following up on our proposed scenario for chondrule formation should consider whether embryo-dust interactions can successfully explain the observed chemical variability of chondrites. In particular, the existence of enstatite, ordinary, and carbonaceous chondrites will need addressing. We speculate that such groups will occur in our scenario due to chondrule formation at a wide range of heliocentric distances. The chemical composition of the embryo atmospheres for example should be complementary to the local nebular environment, which will vary with heliocentric distance, and the longer term cooling pathways and chemical interactions that occur following ejection of chondrules from embryo atmospheres will also receive an imprint from the local Solar nebula.

However, there are also some important challenges that our proposed scenario would need to overcome. These apply in particular not to chondrule origins but to the ease with which the chondrule formation pathway can readily fit with observations of chondrite meteorites. In particular, our model would need to justify that some pathway after dust processing by protoplanetary cores can in fact reproduce observations of chondrite groups. These issues are not settled in the current work. For instance, our model also predicts that planets in the Solar System should each have produced a particular group of chondrules. This prediction may be borne out for enstatite chondrules, which are very similar to Earth and also appear to have formed in the inner Solar System, but it is less clear whether we can so readily link carbonaceous and ordinary chondrules to extant planets, or indeed identify chondrites specifically associated with Mars, and so on.

{Chondrule-matrix complementarity, wherein chondrite matrices are more volatile-rich than chondrules is an especially difficult observation to explain in many scenarios \citep{ebel2}.} It now appears likely that chondrule-matrix complementarity speaks to formation of both from a common chemical reservoir. In our scenario, prior to chondrule formation, nebular gas will be naturally be enriched in volatiles compared to dust grains contained within it.

We have proposed a novel flash-heating mechanism to turn dust into chondrules during their time spent in a protoplanetary atmosphere, but this happens in our scenario after separation from the nebular gas. Volatiles lost from the dust are more likely to be trapped within the protoplanetary atmosphere rather than cycled back into the surrounding nebula. Hence, in our scenario, chondrule-matrix complementarity is implied to have been established earlier -- during the partitioning of volatiles between dust and gas in a warm/hot nebula. Having cycled most of the dust through the chondrule formation mechanism, it would then need to be the case that the nebular temperature drops, condensing much of the remaining gas into fine-grained volatile-rich matrix material and producing chondrite bodies of the observed compositions.

Whether the above is reasonable for the origin of chondrite bodies will require further work to assess. However, we can already point to long-standing evidence of size-sorting of chondrules between H, L, and LL bodies and oxygen isotope evidence of a common source reservoir as suggesting a substantial residence time of chondrules in the nebula after their formation but before incorporation into chondrites bodies. These observations partially relax constraints on chondrule origins that may otherwise require their formation simultaneously with matrix. Again, future work will be needed to establish whether a single core can give rise to the right diversity and amount of material to eventually yield H, L, and LL bodies.

More detailed work should also be done to ascertain whether the pressure-temperature-time pathways of chondrule formation in our model can reproduce the diversity of chondrule textures that are known and their relative proportions. Further work on evaluating the explanatory power of high pressure chondrule formation conditions would also be highly valuable for assessing our scenario.

The implications of chondrule formation via dust accretion to planetary embryos may be profound. It would imply that chondrules and therefore chondrites are not a universal component of all solar systems, but instead only those where  planet formation took place. Furthermore, it implies that chondrule compositions will be closely related to the nature of the formation path of individual planetary embryos and their envelopes.

%{Given that chondritic material is thought to have delivered volatile elements to inner Solar System terrestrial worlds during a phase of late accretion \citep{Braukmüller:2019} (Fig. 1), our results have important implications for planetary habitability.} {Specifically, our model suggests that the materials needed to produce habitable conditions on terrestrial worlds are produced by planets themselves.} {In this view, planetary embryos, especially those formed beyond the habitable zone, may have a final say in whether or not inner solar system terrestrial worlds become habitable.}

%We will now detail the ways in which our model output closely matches chondrule observations, whilst providing important context in the form of summarized results from alternative modeled scenarios for chondrule formation.

%DOES OUR MODEL AGREE WITH THE MASS OF CHONDRULES PRODUCED AND THE PLACES INFERRED FOR THEIR PRODUCTION ACROSS THE NEBULA?

%FOR THE AGES? (i.e., when should chondrules begin and stop forming, according to our model)

%WITH THE NUMBER OF DISTINCT CHONDRULE TYPES? I.E., SINCE H CHONDRITES HAVE H CHONDRULES, DOES THAT IMPLY 3 UNIQUE CORES FOR H, L, AND LL? AND A LOT MORE FOR THE CCS? DOES THIS MAKE SENSE?

\section*{Acknowledgements}
MAD was supported by Tamkeen under the NYU Abu Dhabi Research Institute grant CASS. CRW was supported by the NOMIS Foundation and ETH Zürich, as well as a Junior Research Fellowship from Trinity College (University of Cambridge).

\section*{Data Availability}
The data underlying this article will be shared on reasonable request to the corresponding author.

%%%%%%%%%%%%%%%%%%%%%%%%%%%%%%%%%%%%%%%%%%%%%%%%%%



%\bibliographystyle{mnras}


%\printbibliography

%%%%%%%%%%%%%%%%%%%%%%%%%%%%%%%%%%%%%%%%%%%%%%%%%%

%%%%%%%%%%%%%%%%% APPENDICES %%%%%%%%%%%%%%%%%%%%%

%\section*{APPENDIX}
\newpage

\begin{appendices}

%\begin{figure}
 %   \centering
  %  \includegraphics[width=8cm]{param_space.jpg}
    %\caption{Cooling rates of %experiments found to produce mineral textures consistent with chondrules. Bow shock models occupy only the highest cooling rates in this parameter space. Background image adapted from \citep{Desch:2012}}
   % \label{fig:enter-label}
%\end{figure}

%\begin{figure}
  %  \centering
  %  \includegraphics[width=8cm]{cooling_Schraeder.png}
  %  \caption{"Chondrule cooling rates, K vs. K/hr: > 2023 K from (Yu and Hewins, 1998); 1673–2023 K from (Radomsky and Hewins, 1990; Desch and Connolly, 2002; Hewins et al., 2005; Berlin et al., 2011; Desch et al., 2012); 1173–1673 K from (Weinbruch et al., 2001; Humayun, 2012; Chaumard et al., 2018); 1073–1273 K from (Wick and Jones, 2012); and ~1273 K from (Tachibana et al., 2006, Schrader and Lauretta, 2010). The cooling rate for 503–873 K is from this study." --> schraeder et al 2018}
  %  \label{fig:enter-label}
%\end{figure}

%\begin{figure}
  %  \centering
  %  \includegraphics[width=9.5cm]{Size_chondrules_Freidrich.png}
  %  \caption{...}
   % \label{fig:enter-label}
%\end{figure}

\end{appendices}

\bibliographystyle{mnras} 
\bibliography{arxiv_v1.bbl}

\begin{thebibliography}{}
\makeatletter
\relax
\def\mn@urlcharsother{\let\do\@makeother \do\$\do\&\do\#\do\^\do\_\do\%\do\~}
\def\mn@doi{\begingroup\mn@urlcharsother \@ifnextchar [ {\mn@doi@} {\mn@doi@[]}}
\def\mn@doi@[#1]#2{\def\@tempa{#1}\ifx\@tempa\@empty \href {http://dx.doi.org/#2} {doi:#2}\else \href {http://dx.doi.org/#2} {#1}\fi \endgroup}
\def\mn@eprint#1#2{\mn@eprint@#1:#2::\@nil}
\def\mn@eprint@arXiv#1{\href {http://arxiv.org/abs/#1} {{\tt arXiv:#1}}}
\def\mn@eprint@dblp#1{\href {http://dblp.uni-trier.de/rec/bibtex/#1.xml} {dblp:#1}}
\def\mn@eprint@#1:#2:#3:#4\@nil{\def\@tempa {#1}\def\@tempb {#2}\def\@tempc {#3}\ifx \@tempc \@empty \let \@tempc \@tempb \let \@tempb \@tempa \fi \ifx \@tempb \@empty \def\@tempb {arXiv}\fi \@ifundefined {mn@eprint@\@tempb}{\@tempb:\@tempc}{\expandafter \expandafter \csname mn@eprint@\@tempb\endcsname \expandafter{\@tempc}}}

\bibitem[\protect\citeauthoryear{Alexander, Grossman, Ebel  \& Ciesla}{Alexander et~al.}{2008}]{Alexander:2008}
Alexander C. M.~O.,  Grossman J.~N.,  Ebel D.~S.,   Ciesla F.~J.,  2008, \mn@doi [Science (New York, N.Y.)] {10.1126/science.1156561}, 320, 1617

\bibitem[\protect\citeauthoryear{{Ali-Dib}}{{Ali-Dib}}{2023}]{alidib}
{Ali-Dib} M.,  2023, \mn@doi [\mnras] {10.1093/mnrasl/slad002}, \href {https://ui.adsabs.harvard.edu/abs/2023MNRAS.520L..48A} {520, L48}

\bibitem[\protect\citeauthoryear{{Ali-Dib} \& {Thompson}}{{Ali-Dib} \& {Thompson}}{2020}]{alidibthompson}
{Ali-Dib} M.,  {Thompson} C.,  2020, \mn@doi [\apj] {10.3847/1538-4357/aba521}, \href {https://ui.adsabs.harvard.edu/abs/2020ApJ...900...96A} {900, 96}

\bibitem[\protect\citeauthoryear{Ali-Dib, Cumming  \& Lin}{Ali-Dib et~al.}{2020}]{alidibcumming}
Ali-Dib M.,  Cumming A.,   Lin D. N.~C.,  2020, Monthly Notices of the Royal Astronomical Society

\bibitem[\protect\citeauthoryear{{Ali-Dib}, {Cumming}  \& {Lin}}{{Ali-Dib} et~al.}{2022}]{alidibcoll}
{Ali-Dib} M.,  {Cumming} A.,   {Lin} D. N.~C.,  2022, \mn@doi [\mnras] {10.1093/mnras/stab3008}, \href {https://ui.adsabs.harvard.edu/abs/2022MNRAS.509.1413A} {509, 1413}

\bibitem[\protect\citeauthoryear{{Blum} \& {Wurm}}{{Blum} \& {Wurm}}{2008}]{blum}
{Blum} J.,  {Wurm} G.,  2008, \mn@doi [\araa] {10.1146/annurev.astro.46.060407.145152}, \href {https://ui.adsabs.harvard.edu/abs/2008ARA&A..46...21B} {46, 21}

\bibitem[\protect\citeauthoryear{{Boley}, {Morris}  \& {Desch}}{{Boley} et~al.}{2013}]{pressure1}
{Boley} A.~C.,  {Morris} M.~A.,   {Desch} S.~J.,  2013, \mn@doi [\apj] {10.1088/0004-637X/776/2/101}, \href {https://ui.adsabs.harvard.edu/abs/2013ApJ...776..101B} {776, 101}

\bibitem[\protect\citeauthoryear{Bollard et~al.,}{Bollard et~al.}{2017}]{Bollard:2017}
Bollard J.,  et~al., 2017, \mn@doi [Science Advances] {10.1126/sciadv.1700407}, 3, e1700407

\bibitem[\protect\citeauthoryear{{Bond}, {O'Brien}  \& {Lauretta}}{{Bond} et~al.}{2010}]{chem2}
{Bond} J.~C.,  {O'Brien} D.~P.,   {Lauretta} D.~S.,  2010, \mn@doi [\apj] {10.1088/0004-637X/715/2/1050}, \href {https://ui.adsabs.harvard.edu/abs/2010ApJ...715.1050B} {715, 1050}

\bibitem[\protect\citeauthoryear{{Brett} \& {Sato}}{{Brett} \& {Sato}}{1984}]{fug1}
{Brett} R.,  {Sato} M.,  1984, \mn@doi [\gca] {10.1016/0016-7037(84)90353-3}, \href {https://ui.adsabs.harvard.edu/abs/1984GeCoA..48..111B} {48, 111}

\bibitem[\protect\citeauthoryear{Bridges, Changela, Nayakshin, Starkey  \& Franchi}{Bridges et~al.}{2012}]{Bridges:2012}
Bridges J.,  Changela H.,  Nayakshin S.,  Starkey N.,   Franchi I.,  2012, \mn@doi [Earth and Planetary Science Letters] {10.1016/j.epsl.2012.06.011}, 341, 186

\bibitem[\protect\citeauthoryear{Ciesla}{Ciesla}{2007}]{Ciesla:2007}
Ciesla F.~J.,  2007, \mn@doi [Science (New York, N.Y.)] {10.1126/science.1147273}, 318, 613

\bibitem[\protect\citeauthoryear{Clayton, Mayeda, Goswami  \& Olsen}{Clayton et~al.}{1991}]{Clayton:1991}
Clayton R.~N.,  Mayeda T.~K.,  Goswami J.,   Olsen E.~J.,  1991, \mn@doi [Geochimica et Cosmochimica Acta] {10.1016/0016-7037(91)90107-g}, 55, 2317

\bibitem[\protect\citeauthoryear{Connelly, Bizzarro, Krot, Nordlund, Wielandt  \& Ivanova}{Connelly et~al.}{2012}]{Connelly:2012}
Connelly J.~N.,  Bizzarro M.,  Krot A.~N.,  Nordlund {\AA}.,  Wielandt D.,   Ivanova M.~A.,  2012, \mn@doi [Science] {10.1126/science.1226919}, 338, 651

\bibitem[\protect\citeauthoryear{Connolly \& Jones}{Connolly \& Jones}{2016a}]{Connolly:2016}
Connolly H.~C.,  Jones R.~H.,  2016a, \mn@doi [Journal of Geophysical Research: Planets] {10.1002/2016je005113}, 121, 1885

\bibitem[\protect\citeauthoryear{{Connolly} \& {Jones}}{{Connolly} \& {Jones}}{2016b}]{connolly}
{Connolly} H.~C.,  {Jones} R.~H.,  2016b, \mn@doi [Journal of Geophysical Research (Planets)] {10.1002/2016JE005113}, \href {https://ui.adsabs.harvard.edu/abs/2016JGRE..121.1885C} {121, 1885}

\bibitem[\protect\citeauthoryear{Deienno, Nesvorný, Clement, Bottke, Izidoro  \& Walsh}{Deienno et~al.}{2024}]{Deienno:2024}
Deienno R.,  Nesvorný D.,  Clement M.~S.,  Bottke W.~F.,  Izidoro A.,   Walsh K.~J.,  2024, \mn@doi [The Planetary Science Journal] {10.3847/psj/ad3a68}, 5, 110

\bibitem[\protect\citeauthoryear{Desch, Morris, Connolly  \& Boss}{Desch et~al.}{2010}]{Desch:2010}
Desch S.~J.,  Morris M.~A.,  Connolly H.~C.,   Boss A.~P.,  2010, \mn@doi [arXiv] {10.48550/arxiv.1011.3483}

\bibitem[\protect\citeauthoryear{Desch, Morris, Connolly  \& Boss}{Desch et~al.}{2012}]{Desch:2012}
Desch S.~J.,  Morris M.~A.,  Connolly H.~C.,   Boss A.~P.,  2012, \mn@doi [Meteoritics \& Planetary Science] {10.1111/j.1945-5100.2012.01357.x}, 47, 1139

\bibitem[\protect\citeauthoryear{{Ebel} \& {Grossman}}{{Ebel} \& {Grossman}}{2000}]{ebel}
{Ebel} D.~S.,  {Grossman} L.,  2000, \mn@doi [\gca] {10.1016/S0016-7037(99)00284-7}, \href {https://ui.adsabs.harvard.edu/abs/2000GeCoA..64..339E} {64, 339}

\bibitem[\protect\citeauthoryear{Ebel, Alexander  \& Libourel}{Ebel et~al.}{2023}]{Ebel:2023}
Ebel D.~S.,  Alexander C. M.~O.,   Libourel G.,  2023, \mn@doi [arXiv] {10.48550/arxiv.2306.12664}

\bibitem[\protect\citeauthoryear{{Freedman}, {Lustig-Yaeger}, {Fortney}, {Lupu}, {Marley}  \& {Lodders}}{{Freedman} et~al.}{2014}]{freedman}
{Freedman} R.~S.,  {Lustig-Yaeger} J.,  {Fortney} J.~J.,  {Lupu} R.~E.,  {Marley} M.~S.,   {Lodders} K.,  2014, \mn@doi [\apjs] {10.1088/0067-0049/214/2/25}, \href {https://ui.adsabs.harvard.edu/abs/2014ApJS..214...25F} {214, 25}

\bibitem[\protect\citeauthoryear{Friedrich, Chen, Giordano, Matalka, Strasser, Tamucci, Rivers  \& Ebel}{Friedrich et~al.}{2022}]{Friedrich:2022}
Friedrich J.~M.,  Chen M.~M.,  Giordano S.~A.,  Matalka O.~K.,  Strasser J.~W.,  Tamucci K.~A.,  Rivers M.~L.,   Ebel D.~S.,  2022, \mn@doi [Microscopy Research and Technique] {10.1002/jemt.24043}, 85, 1814

\bibitem[\protect\citeauthoryear{Galy, Young, Ash  \& O'Nions}{Galy et~al.}{2000}]{Galy:2000}
Galy A.,  Young E.~D.,  Ash R.~D.,   O'Nions R.~K.,  2000, \mn@doi [Science] {10.1126/science.290.5497.1751}, 290, 1751

\bibitem[\protect\citeauthoryear{Jr. \& Love}{Jr. \& Love}{1998}]{Connolly:1998}
Jr. H. C.~C.,  Love S.~G.,  1998, \mn@doi [Science] {10.1126/science.280.5360.62}, 280, 62

\bibitem[\protect\citeauthoryear{Krot, Amelin, Cassen  \& Meibom}{Krot et~al.}{2005}]{Krot:2005}
Krot A.~N.,  Amelin Y.,  Cassen P.,   Meibom A.,  2005, \mn@doi [Nature] {10.1038/nature03830}, 436, 989

\bibitem[\protect\citeauthoryear{{Lambrechts} \& {Johansen}}{{Lambrechts} \& {Johansen}}{2014}]{lamb}
{Lambrechts} M.,  {Johansen} A.,  2014, \mn@doi [\aap] {10.1051/0004-6361/201424343}, \href {https://ui.adsabs.harvard.edu/abs/2014A&A...572A.107L} {572, A107}

\bibitem[\protect\citeauthoryear{Lichtenberg, Golabek, Dullemond, Schönbächler, Gerya  \& Meyer}{Lichtenberg et~al.}{2018}]{Lichtenberg:2018}
Lichtenberg T.,  Golabek G.~J.,  Dullemond C.~P.,  Schönbächler M.,  Gerya T.~V.,   Meyer M.~R.,  2018, \mn@doi [Icarus] {10.1016/j.icarus.2017.11.004}, 302, 27

\bibitem[\protect\citeauthoryear{{Madhusudhan}}{{Madhusudhan}}{2012}]{madhu}
{Madhusudhan} N.,  2012, \mn@doi [\apj] {10.1088/0004-637X/758/1/36}, \href {https://ui.adsabs.harvard.edu/abs/2012ApJ...758...36M} {758, 36}

\bibitem[\protect\citeauthoryear{{Mann}, {Boley}  \& {Morris}}{{Mann} et~al.}{2016}]{pressure2}
{Mann} C.~R.,  {Boley} A.~C.,   {Morris} M.~A.,  2016, \mn@doi [\apj] {10.3847/0004-637X/818/2/103}, \href {https://ui.adsabs.harvard.edu/abs/2016ApJ...818..103M} {818, 103}

\bibitem[\protect\citeauthoryear{Merrill}{Merrill}{1920}]{Merill:1920}
Merrill G.~P.,  1920, \mn@doi [Proceedings of the National Academy of Sciences of the United States of America] {10.1073/pnas.6.8.449}, 6, 449

\bibitem[\protect\citeauthoryear{{Morbidelli}, {Kleine}  \& {Nimmo}}{{Morbidelli} et~al.}{2025}]{morby2025}
{Morbidelli} A.,  {Kleine} T.,   {Nimmo} F.,  2025, \mn@doi [Earth and Planetary Science Letters] {10.1016/j.epsl.2024.119120}, \href {https://ui.adsabs.harvard.edu/abs/2025E&PSL.65019120M} {650, 119120}

\bibitem[\protect\citeauthoryear{{Ormel}}{{Ormel}}{2014}]{ormel}
{Ormel} C.~W.,  2014, \mn@doi [\apjl] {10.1088/2041-8205/789/1/L18}, \href {https://ui.adsabs.harvard.edu/abs/2014ApJ...789L..18O} {789, L18}

\bibitem[\protect\citeauthoryear{{Palme}, {Hezel}  \& {Ebel}}{{Palme} et~al.}{2015}]{ebel2}
{Palme} H.,  {Hezel} D.~C.,   {Ebel} D.~S.,  2015, \mn@doi [Earth and Planetary Science Letters] {10.1016/j.epsl.2014.11.033}, \href {https://ui.adsabs.harvard.edu/abs/2015E&PSL.411...11P} {411, 11}

\bibitem[\protect\citeauthoryear{{Righter} \& {Neff}}{{Righter} \& {Neff}}{2007}]{fug3}
{Righter} K.,  {Neff} K.~E.,  2007, \mn@doi [Polar Science] {10.1016/j.polar.2007.04.002}, \href {https://ui.adsabs.harvard.edu/abs/2007PolSc...1...25R} {1, 25}

\bibitem[\protect\citeauthoryear{Schaefer \& Fegley}{Schaefer \& Fegley}{2010}]{fegley}
Schaefer L.,  Fegley B.,  2010, \mn@doi [Icarus] {https://doi.org/10.1016/j.icarus.2010.01.026}, 208, 438

\bibitem[\protect\citeauthoryear{Schrader, Fu, Desch  \& Davidson}{Schrader et~al.}{2018}]{Schraeder:2018}
Schrader D.~L.,  Fu R.~R.,  Desch S.~J.,   Davidson J.,  2018, \mn@doi [Earth and Planetary Science Letters] {10.1016/j.epsl.2018.09.030}, 504, 30

\bibitem[\protect\citeauthoryear{Shu, Shang, Glassgold  \& Lee}{Shu et~al.}{1997}]{Shu:1997}
Shu F.~H.,  Shang H.,  Glassgold A.~E.,   Lee T.,  1997, \mn@doi [Science] {10.1126/science.277.5331.1475}, 277, 1475

\bibitem[\protect\citeauthoryear{{Stewart}, {Lock}, {Carter}, {Davies}, {Petaev}  \& {Jacobsen}}{{Stewart} et~al.}{2025}]{pressure3}
{Stewart} S.~T.,  {Lock} S.~J.,  {Carter} P.~J.,  {Davies} E.~J.,  {Petaev} M.~I.,   {Jacobsen} S.~B.,  2025, \mn@doi [arXiv e-prints] {10.48550/arXiv.2503.05636}, \href {https://ui.adsabs.harvard.edu/abs/2025arXiv250305636S} {p. arXiv:2503.05636}

\bibitem[\protect\citeauthoryear{{Tenner}, {Ushikubo}, {Nakashima}, {Schrader}, {Weisberg}, {Kimura}  \& {Kita}}{{Tenner} et~al.}{2018}]{fug2}
{Tenner} T.~J.,  {Ushikubo} T.,  {Nakashima} D.,  {Schrader} D.~L.,  {Weisberg} M.~K.,  {Kimura} M.,   {Kita} N.~T.,  2018, in {Russell} S.~S.,  {Connolly} Jr. H.~C.,   {Krot} A.~N.,  eds, , Chondrules: Records of Protoplanetary Disk Processes.
pp 196--246, \mn@doi{10.1017/9781108284073.008}

\bibitem[\protect\citeauthoryear{Yap \& Tissot}{Yap \& Tissot}{2023}]{Yap:2023}
Yap T.~E.,  Tissot F.~L.,  2023, \mn@doi [Icarus] {10.1016/j.icarus.2023.115680}, 405, 115680

\makeatother
\end{thebibliography}

%%%%%%%%%%%%%%%%%%%%%%%%%%%%%%%%%%%%%%%%%%%%%%%%%%

% Don't change these lines
\bsp	% typesetting comment
\label{lastpage}
\end{document}